\documentclass{article}
\usepackage[small,it]{caption}
\usepackage[utf8]{inputenc}
\usepackage[small,compact]{titlesec}

\usepackage{graphicx,color}
\usepackage[cmex10]{amsmath}
\usepackage{amsfonts}
\usepackage{float}
\usepackage{algorithmic}
\usepackage{algorithm}
\usepackage[caption=false,font=footnotesize]{subfig}
\usepackage{url}
\usepackage{booktabs}
\usepackage{enumerate}
\newcommand{\FF}{\mathbb{F}}
\newcommand\bv[1]{{\bf #1}}
\newcommand\co[1]{}
\newcommand\rt{\!\!\vdash\!\!}
\newtheorem{lemma}{Lemma}
\newtheorem{example}{Example}
\newtheorem{definition}{Definition}
\newtheorem{remark}{Remark}
\newtheorem{assumption}{Assumption}
\newcommand{\comment}[1]{}
\usepackage[margin=1in]{geometry}

\title{In-Network Redundancy Generation for Opportunistic Speedup of Data Backup}

\author{{\rm Lluis Pamies-Juarez, Anwitaman Datta and Fr\'{e}d\'{e}rique Oggier}\\
Nanyang Technological University\footnote{
The research of L. Pamies-Juarez and F. Oggier is supported by the Singapore
National Research Foundation under Research Grant NRF-CRP2-2007-03. A. Datta's
work has been supported by A*Star TSRP grant number 102-158-0038.}, Singapore}

\begin{document}

\maketitle

\begin{abstract}
Erasure coding is a storage-efficient alternative to replication for achieving
reliable data backup in distributed storage systems. During the storage
process, traditional erasure codes require a unique source node to create and
upload all the redundant data to the different storage nodes. However, such a
source node may have limited communication and computation capabilities, which
constrain the storage process throughput. Moreover, the source node and the
different storage nodes might not be able to send and receive data
simultaneously -- e.g., nodes might be busy in a datacenter setting, or simply
be offline in a peer-to-peer setting -- which can further threaten the efficacy
of the overall storage process. In this paper we propose an ``in-network''
redundancy generation process which distributes the data insertion load among
the source and storage nodes by allowing the storage nodes to generate new
redundant data by exchanging partial information among themselves, improving
the throughput of the storage process. The process is carried out
asynchronously, utilizing spare bandwidth and computing resources from the
storage nodes. The proposed approach leverages on the local repairability
property of newly proposed erasure codes tailor made for the needs of
distributed storage systems. We analytically show that the performance of this
technique relies on an efficient usage of the spare node resources, and we
derive a set of scheduling algorithms to maximize the same.  We experimentally
show, using availability traces from real peer-to-peer applications as well as
Google data center availability and workload traces, that our algorithms can,
depending on the environment characteristics, increase the throughput of the
storage process significantly (up to 90\% in data centers, and 60\% in
peer-to-peer settings) with respect to the classical naive data insertion
approach.
\end{abstract}


\section{Introduction}

There is a continuous and rapid global growth in data storage needs. Archival
and backup storage form a specific niche of importance to both businesses and
individuals. A recent market analysis from
IDC\footnote{\url{http://www.idc.com/getdoc.jsp?containerId=230762}} stated
that the global revenue of the data archival business is expected to reach
\$6.5 billion in 2015. The necessity to cost-effectively scale-up data backup
systems to meet this ever growing storage demand poses a challenge to storage
systems designers.

When a large volume of data is involved, deploying a networked distributed
storage system becomes essential, since a single storage node cannot scale.
Furthermore, distribution provides opportunities for fault tolerance and
parallelized I/O. Examples of such distributed storage systems are readily
found in datacenter environments, including distributed file systems such as
GFS~\cite{googlefs} or HDFS~\cite{hadoopfs}, distributed key-value stores like
Dynamo~\cite{dynamo} or Cassandra~\cite{cassandra}, for storing huge volume of
scientific or multimedia content \cite{Sector,CineGrid} as well as in ad-hoc
end user resource based peer-to-peer (P2P) settings such as
OceanStore~\cite{oceanstore} and friend-to-friend (F2F) storage
systems~\cite{f2favail,f2fstorage}, a special kind of peer-to-peer systems
often considered particularly suitable for personal data backup.

An important design aspect in distributed storage systems is redundancy
management. Data replication provides a simple way to achieve high
fault-tolerance, while erasure codes such as Reed-Solomon
codes~\cite{reedsolomon} are more sophisticated alternatives, capable of
significantly reducing the data storage footprint for different levels of
fault-tolerance~\cite{highavail,ecreplication,codingvsrepl,Shum,collabregenerating}.
Various trade-offs in adopting erasure codes in storage systems, such as
storage overhead \& fault-tolerance, access frequency \& decoding overheads,
but also due to the need of replenishment of lost redundancy, repair bandwidth
\& repair time after failures, have been studied in the literature, revealing
in particular that erasure codes are particularly suited for backup and
archival storage, where data access is infrequent, and hence the effects of
decoding are marginal. Consequently, many modern data center oriented storage
and file systems such as Microsoft's Azure \cite{azureec}, HDFS \cite{hadoopec}
and the next generation of Google's file system (Colossus) have incorporated
erasure codes to enhance the systems' storage efficiency. However, a relatively
unexplored aspect of the usage of erasure codes in storage systems --both in
data center as well as peer-to-peer environments-- is that of the time required
for inserting the data (along with the necessary redundancy for fault-tolerance
and durability).

When using replication, a source node aiming to store new data can upload one replica of this data to the
first storage node, which can concurrently forward the same data to a second storage node, and so on. By
applying such a pipelining mechanism, the load for redundancy insertion can be shared, and the source node does not need to upload any redundant information itself. Replication thus naturally supports ``in-network''
generation of redundancy, that is, generation of new redundancy within the network, through data exchange among storage
nodes, which in turn leads to fast insertion of data. In contrast, in erasure encoded systems, the source
node\footnote{In this paper we call the node which has a full copy of the data, and is primarily responsible
for inserting the data redundantly in the rest of the system as the `source node', irrespectively of the origin and
ownership of the data.} is the only one responsible for computing and uploading all the encoded redundant data to the
corresponding storage nodes. This has traditionally been the case, because typically the coding process is
centralized in nature, and hence all encoded data fragments are generated at a single node, which then also bears the
load of inserting the encoded fragments at other storage nodes. The amount of data the source node uploads is then
considerably larger, including the data object and its corresponding redundancy, resulting in lower
data insertion throughput. Insertion throughput may further be exacerbated when the source node and the set
of storage nodes have additional (mismatched) temporal constraints on resources availability, in which case in-network
redundancy generation can provide partial mitigation.

We elaborate the effect of temporal resource (un)availability issues with two distinct example scenarios, which we also
use later in our experiments to determine how (much) in-network redundancy generation may improve the data insertion
throughput:
\begin{enumerate}[(i)]
\item In datacenters, storage nodes might be used for computation processes which require efficient access to local
disks. Since backup processes consume large amounts of local disk I/O, system administrators might want to avoid backup
transfers while nodes are executing I/O intensive tasks -- e.g., Mapreduce tasks.
\item In peer-to-peer settings, users exchange some of their spare disk resources in order to realize a
collaborative data backup service. Such resource sharing may furthermore be driven by other constraints such as trust or
friendship. However, desirable users may be online at different times of the day, complicating the data insertion
process.
\end{enumerate}
In both cases the insertion of new redundancy by the source node is restricted to the periods when the availability
windows of the source overlap that of the storage nodes.

Unlike replication, where in-network redundancy generation is achieved trivially, traditional erasure codes are not easily amenable. Our solution is based on new classes of erasure codes designed for better repair
efficiency, obtained through local repairability, that is the property of being able to repair a failure by contacting
only a small number of live nodes (e.g. \cite{OD,local}). In particular, to provide a concrete instance that
will be used later for experimentation, we will focus on a novel family of erasure codes called Self-Repairing Codes
(SRC)~\cite{OD}, whose salient property is that the encoded data stored at each node can be easily
regenerated by using information from typically only two live storage nodes. As we will show,
local repairability is the key property to achieve in-network redundancy generation. However, SRC have
strict constraints on how storage nodes can combine their data to generate content for other nodes, which, along with
the temporal availability constraints of nodes, complicate the design of efficient in-network redundancy generation.

In this paper we provide an analytical framework to define valid transfer
schedules for the in-network redundancy generation using SRC. We then
identify the main requirements that valid scheduling algorithms must satisfy,
and show that determining optimal schedules is computationally intractable.
Accordingly, we explore several heuristic algorithm implementations which aim
at maximizing the utilization of the spare resources of the storage nodes,
thus improving the backup throughput. We determine the efficacy of our
approach in different environments by using Google data center availability
and workload traces, and availability traces from real friend-to-friend
(F2F)~\cite{f2favail} and peer-to-peer (P2P) applications~\cite{globalkad}.
Our results show that for data center traces the algorithms proposed
for the in-network redundancy generation can increase the throughput of the
storage process up to 90\% as compared to classical naive storage approaches,
while only requiring 50\% more of the network resources, but using it only
when it is otherwise unused. For P2P/F2F traces the throughout can increase
up to 60\% while requiring 38\% additional (but again, spare) network
resources.

The main contributions of this paper can be summarized as follows:
\begin{enumerate}[(i)]
\item We introduce the concept of \emph{in-network redundancy
generation} for reducing data insertion latency in erasure code
based storage systems, and demonstrate its feasibility
using one specific instance of locally repairable code, namely
self-repairing codes.
\item We define an \emph{analytical framework} to explore valid data transfer schedules where the
in-network redundancy generation process maximizes the use of the available network resources.

\item We show that besides requiring a node availability prediction, determining the optimal data insertion schedule is computationally intractable.
\item We propose a set of \emph{heuristics} for efficient in-network redundancy generation.
\item We determine the efficacy of in-network redundancy
generation in diverse distributed storage environments using real
workload and availability trace driven simulations.
\end{enumerate}

The rest of the paper is organized as follows. In
Section~\ref{s:background} we provide the background on erasure
codes and self-repairing codes, a particular instance of locally repairable codes.
In Section~\ref{s:redgen} we define our in-network
redundancy generation process and in Section~\ref{s:scheduling} we
show how to schedule the redundancy generation to speedup the
insertion of new data. Due to the complexity of determining optimal
schedules, we propose in Section~\ref{s:algorithm} several heuristic
scheduling algorithms, which are evaluated in
Section~\ref{s:eval} using real availability traces.
Finally in Section~\ref{s:conclusions} we state our conclusions and
further research directions.

\section{Background}
\label{s:background}

The P2P research community has long studied the applicability of erasure codes
in low availability environments with limited storage
capacity~\cite{oceanstore,codingvsrepl}. The growing interest in applying
erasure codes in data centers is more recent, and aims at reducing the storage
costs~\cite{hadoopec,ecgoogle,diskreduce,dress,Calder}. It further triggered a
line of research around repairability of storage systems, i.e., how lost
redundancy can be replenished, which includes the application of network coding
\cite{collabregenerating,Shum} to carry out the repair process in a
decentralized manner, designing novel codes with inherent local repairability
\cite{OD,local}, as well as engineering solutions such as applying multiple
levels of encoding on the same \cite{hierarchicalcodes} or even across
different data objects \cite{apsys}.

While fault-tolerance, storage overhead, repairability, but also I/O and
bandwidth are well recognized critical bottlenecks for the storage of huge
amounts of data, existing literature does not explore yet how data insertion
can be optimized in the context of erasure codes based storage. This work
leverages on one of these recent approaches on distributed storage systems
repairability, namely, novel codes with local repairability, in order to
improve the very process of data insertion.

In the following, we provide some background on erasure codes as classically
used for distributed storage, as well as on one specific instance of locally
repairable codes, Self-Repairing Codes (SRC), which we use in the rest of the
paper to demonstrate the feasibility and quantify the benefits of in-network
redundancy generation in erasure code based distributed storage systems.

\subsection{Erasure codes for distributed storage}

A classical $\langle n,k \rangle$ erasure code allows to redundantly encode an
object of size $M$ into $n$ redundant fragments of size $M/k$, each to be
stored in a different storage node. The data storage overhead (or redundancy
factor) is then given by $n/k$, and the stored object can be reconstructed by
downloading an amount of data equal to $M$, from $k$ or more different nodes
out of $n$.

One of the main drawbacks of using classical erasure codes for storage is that
redundant fragments can only be generated by applying coding operations on the
original data. The generation of new redundancy is then restricted to nodes
that possess the original object (or a copy), namely: the source node, storage
nodes that previously reconstructed the original object, or possibly were
storing a copy (as is the case in a hybrid model where a full copy of the
object is kept, together with encoded fragments). When the original raw object
is not available, repairing a single node failure consequently entails
downloading an amount of information equivalent to the size of the original
object, causing a significant communication overhead.

In order to mitigate this communication overhead, a new family of erasure codes
called Regenerating Codes (see e.g.  \cite{collabregenerating,Shum} and
reference therein) was recently designed by adopting ideas from network
coding~\cite{netcod}, a popular mechanism deployed to improve the throughput
utilization of a given network topology. The main advantage of Regenerating
Codes is that new redundant fragments can be generated by downloading an amount
of data $\beta$ from $d$ other redundant fragments, where $d\geq k$, and
$\beta\leq M/k$. Unlike in classical erasure codes, Regenerating Codes can thus
repair missing fragments by downloading only an amount of data equal to
$d\beta$, where usually $d\beta\ll M$. However, the maximum communication
savings occur for large values of $d$, in which cases however, the chance to
find $d$ available nodes might be very low, limiting the practicality of such
codes.

\subsection{Homomorphic Self-Repairing Codes (HSRC)}

Self-Repairing Codes (SRC)~\cite{OD} are a new family of erasure codes designed to minimize the maintenance
overhead by reducing the number of nodes $d$ required to be contacted to recreate lost fragments. A specific family of
SRC, named Homomorphic Self-Repairing codes (HSRC), has the property that two encoded
fragments can be xored for such a regeneration, i.e. $d=2$, as long as not more than half of the nodes have
failed. This locally repairable property makes HSRC suitable for the in-network redundancy
generation since partial redundant data stored in two different nodes can be used to generate data for a third node,
without requiring the intervention of the source node. However, as we will show below, the pairs of nodes used for that
purpose cannot be arbitrary chosen.

Let us recall briefly the construction of HSRC. We denote finite fields by
$\mathbb F$. The cardinality of $\mathbb F$ is given by its index, that is,
$\mathbb F_2$ is the binary field with two elements (the two bits 0 and 1), and
$\mathbb F_q$ is the finite field with $q$ elements. If $q=2^m$, for some
positive integer $m$, we can fix a $\mathbb F_2\text{-basis}$ of $\mathbb F_q$
and represent an element $\bv x\in\mathbb F_{2^m}$ using an $m$-dimensional
vector $\bv x=(x_1,\dots,x_m)$ where $x_i\in\mathbb F_{2}$, $i=1,\dots,m$.

Let $\bv o$ be the object to be stored over a set of $n$ nodes, which is represented as a data vector of size $k\times m$ bits, $k\leq m$, i.e.:
$$
\bv o=(o_1,\dots,o_k),~~o_i\in\mathbb F_{2^m}.
$$
Given these $k$ original elements, the $n$ redundant fragments are obtained by evaluating the polynomial

\begin{equation}
p(X)=\sum_{i=1}^{k} o_i X^{2^{(i-1)}} \in \mathbb F_{2^m}[X]
\label{e:poly}
\end{equation}
in $n$ non-zero values $\alpha_1,\dots,\alpha_n$ of $\mathbb F_{2^m}$, yielding the
redundant vector $\bv r$ of size $n\times m$ bits, i.e.:
$$
\bv r=(r_1,\dots,r_n),~~r_i=p(\alpha_i)\in\mathbb F_{2^m}.
$$
In particular we need the code parameters $\langle n,k \rangle$ to satisfy $$1<k<n\leq2^m-1.$$

\section{HSRC Redundancy Generation}
\label{s:redgen}

The main important property of HSRC is its homomorphic property. From~\cite{OD} we have that:
\begin{lemma}
Let $a,b\in\mathbb F_{2^m}$ and let $p(X)$ be the polynomial defined in (\ref{e:poly}), then $p(a+b) = p(a) + p(b)$.
\label{l:homo}
\end{lemma}
This implies that we can generate a redundant element $r_k=p(\alpha_k)$ from
$r_i=p(\alpha_i)$  and  $r_j=p(\alpha_j)$ if and only if
$\alpha_k=\alpha_i+\alpha_j$.  This homomorphic property is one way of
obtaining local repair by contacting only $d=2$ nodes.

\begin{example}\rm\label{ex:k3n7_1}
Consider a $\langle n\!\!=\!\!7,k\!\!=\!\!3\rangle$ HSRC and an object $\bv o=(o_1,o_2,o_3)$ of size $3\times 4$
bits, where $o_i\in\mathbb F_{2^4}$, $i=1,2,3$. We write $o_1=(o_{11},o_{12},o_{13},o_{14})$,
$o_2=(o_{21},o_{22},o_{23},o_{24})$, $o_3=(o_{31},o_{32},o_{33},o_{34})$, from which we compute
$p(X)=\sum_{i=1}^ko_iX^{2^{i-1}}$. We evaluate $p(X)$ in $n=7$ values of $\FF_{2^4}$, represented in vector form as
$\alpha_1=(1,0,0,0)$, $\alpha_2=(0,1,0,0)$, $\alpha_3=(1,1,0,0)$, $\alpha_4=(0,0,1,0)$, $\alpha_5=(1,0,1,0)$, $\alpha_6=(0,1,1,0)$,
$\alpha_7=(1,1,1,0)$, yielding:
\begin{eqnarray*}
r_1=& p(\alpha_1)=&(o_{11}+o_{21}+o_{31},~o_{12}+o_{22}+o_{32},~o_{13}+o_{23}+o_{33}),\\
r_2=& p(\alpha_2)=&(o_{14}+o_{23}+o_{31}+o_{34},o_{11}+o_{14}+o_{23}+o_{24}+o_{31}+o_{32}+\\&&o_{34}+o_{12}+o_{21}+o_{24}+o_{232}+o_{33},o_{13}+o_{22}+o_{33}+o_{34}),\\
r_3=& p(\alpha_3)=&(o_{11}+o_{21}+o_{31}+o_{14}+o_{23}+o_{31}+o_{34},o_{12}+o_{22}+o_{11}+o_{14}+o_{23}+o_{24}+\\&&o_{31}+o_{34}+o_{13}+o_{23}+o_{12}+o_{21}+o_{24}+o_{32},o_{14}+o_{24}+o_{13}+o_{22}+o_{33}),\\
r_4=& p(\alpha_4)=&(o_{13}+o_{24}+o_{21}+o_{31}+o_{33},o_{13}+o_{14}+o_{21}+o_{22}+o_{24}+o_{32}+o_{33}+o_{34}+\\
          && o_{11}+o_{14}+o_{22}+o_{23}+o_{31}+o_{33}+o_{34},o_{12}+o_{23}+o_{24}+o_{32}+o_{34}),\\
r_5=& p(\alpha_5)=&(o_{11}+o_{13}+o_{24}+o_{33},o_{12}+o_{13}+o_{14}+o_{21}+o_{24}+o_{33}+o_{34}+\\
          && o_{11}+o_{13}+o_{14}+o_{22}+o_{31}+o_{34},o_{12}+o_{14}+o_{23}+o_{32}),\\
r_6=& p(\alpha_6)=&(o_{14}+o_{23}+o_{34}+o_{13}+o_{24}+o_{21}+o_{33},o_{11}+o_{23}+o_{31}+o_{13}+o_{21}+o_{22}+o_{33}+o_{12}+o_{21}+\\&&o_{24}+o_{32}+o_{11}+o_{14}+o_{22}+o_{23}+o_{31}+o_{34},o_{13}+o_{22}+o_{33}+o_{12}+o_{23}+o_{24}+o_{32}),\\
r_7=& p(\alpha_7)=&(o_{11}+o_{14}+o_{23}+o_{34}+o_{13}+o_{24}+o_{31}+o_{33},o_{11}+o_{23}+o_{31}+o_{13}+o_{21}+\\
          && o_{12}+o_{33}+o_{32}+o_{12}+o_{21}+o_{24}+o_{32}+o_{11}+o_{14}+o_{22}+o_{33}+o_{31}+\\
          &&o_{34}+o_{13},o_{13}+o_{22}+o_{33}+o_{12}+o_{23}+o_{34}+o_{32}+o_{14})\\
\end{eqnarray*}
with corresponding redundant vector
\[
{\bf r}=(r_1,\ldots,r_7).
\]

We can check that
\begin{align*}
p(\alpha_7)&=p(\alpha_1)+p(\alpha_6)=p(\alpha_2)+p(\alpha_5)=p(\alpha_3)+p(\alpha_4),\\
p(\alpha_6)&=p(\alpha_1)+p(\alpha_7)=p(\alpha_2)+p(\alpha_4)=p(\alpha_3)+p(\alpha_5),
\end{align*}
which illustrates the local repairability of the code.  Note that we have not
used the vector $\alpha_8=(0,0,0,1)$ here, which would have resulted in a
longer code $n>7$.
\end{example}
We now discuss how HSRC operate in two different scenarios: (i) when the source
introduces data in the system, and (ii) during the in-network redundancy
generation.

\subsection{Source Redundancy Generation}

The homomorphic property described in Lemma~\ref{l:homo} has been introduced to repair node failures, though it can
similarly serve to generate redundancy from the source.  Recall from Lemma~\ref{l:homo} that $p(a+b)=p(a)+p(b)$, where
both $a,b$ can be seen as $m$-dimensional binary vectors, by fixing a $\FF_2$-basis of $\FF_{2^m}$. Let us denote this
basis by $\{b_1,\ldots,b_m\}$. Thus $a$ can be written as $a=\sum_{i=1}^ma_ib_i$, $a_i\in\FF_2$, and by virtue of the
homomorphic property, we get that
\[
p(a)=p\left(\sum_{i=1}^m a_ib_i\right)=\sum_{i=1}^m a_ip(b_i).
\]
This means that the source only needs to compute $p(b_1),\ldots,p(b_m)$ for a given basis $\{b_1,\ldots,b_m\}$, after
which all the other encoded fragments are obtained by xoring pairs of elements in $\{p(b_1),\ldots,p(b_m)\}$. Thus, when
using an $\langle n,k\rangle$ HSRC, the source computes $k$ ($k\leq m$) encoded fragments
$r_1=p(\alpha_1),\ldots,r_k=p(\alpha_k)$, where $\alpha_1,\ldots,\alpha_k$ are linearly independent, for example,
$\{\alpha_1,\ldots,\alpha_k\} \subset \{b_1,\ldots,b_m\}$, and then performs the corresponding xoring. The source then
injects the $n$ encoded fragments in the network.

\begin{example}\rm\label{ex:basis}
In Example \ref{ex:k3n7_1}, we have $k=3 \leq m=4$, and a natural $\FF_2$-basis for $\FF_{2^4}$ is $b_1=(1,0,0,0)$,
$b_2=(0,1,0,0)$, $b_3=(0,0,1,0)$, $b_4=(0,0,0,1)$.  The source can generate redundancy by first computing
$r_1=p(\alpha_1)=p(b_1)$, $r_2=p(\alpha_2)=p(b_2)$, $r_4=p(\alpha_4)=p(b_3)$, then $r_3=p(\alpha_3)=p(\alpha_1)+p(\alpha_2)$,
$r_5=p(\alpha_5)=p(\alpha_1)+p(\alpha_4)$, $r_6=p(\alpha_6)=p(\alpha_2)+p(\alpha_4)$ and
$r_7=p(\alpha_7)=p(\alpha_1)+p(\alpha_6)$. The $n=7$ encoded fragments are then ready to be sent over the network.
Further notice that the set $B=\{p(\alpha_1),p(\alpha_2),p(\alpha_4)\}$ can be seen as a basis for the set of redundant
fragments, since they are linearly independent, and can be combined to generate every redundant fragment.
\end{example}

\subsection{In-Network Redundancy Generation}

Let us now consider the case where the source might not inject the whole set of $n$ encoded fragments, but only a
subset $\{ r_i,~i \in I \subset \{1,\ldots,n\}\}$ of the encoded fragments.  We use the triplet notation $(i,j)\rt k$ to
represent the possibility to generate the element $r_k$ by xoring $r_i$ and $r_j$, $r_k=r_i+r_j$. Note that due to the commutativity property of the additive operator, triplets $(i,j)\rt k$ and $(j,i)\rt k$ can be indistinguishably used to
denote the same redundancy generation process. We denote by $\mathcal C$ the set with all the feasible repair
triplets from a set of $n$ redundant elements. Finally, let us define the following two sets:

\begin{definition}[out-creation set]
Let $O(i)$ be the set of all the possible $(i,j)\rt k$ triplets where fragment $r_i$ is used to generate some other fragment:
$$O(i) = \left\{ (i,j)\rt k \mid j=1,\ldots,n,j\neq i,k~\mbox{s.t.}~r_k=r_i+r_j\right\}.$$
\label{d:outset}
\end{definition}

\begin{definition}[in-creation set]
Let $I(k)$ be the set of all the possible $(i,j)\rt k$ triplets that can be used to create $r_k$:
$$I(k) = \left\{ (i,j)\rt k \mid r_k=r_i+r_j;~i,j=1,\dots,n\right\}.$$
\label{d:inset}
\end{definition}

Finally, given a number of redundant elements $n=2^t-1$, for any positive integer $t\leq m$, we have from~\cite{OD} that:
\begin{align}
|O(i)| &= n-1 \\
|I(k)| &= (n-1)/2.
\end{align}

\begin{example}\rm
In Example \ref{ex:k3n7_1}, we have that
\begin{align*}
O(1)&=\left\{(1,3)\rt2,(1,2)\rt3,(1,5)\rt4,(1,4)\rt5,(1,7)\rt6,(1,6)\rt7\right\}, \\
I(7)&=\left\{(1,6)\rt7,(2,5)\rt7,(3,4)\rt7\right\}.
\end{align*}
\end{example}

\subsection{HSRC: Practical Implementation}
\label{s:practical}

Previously we detailed how to encode a data vector $\bv o$ of size $k\times m$ bits into a redundant vector $\bv r$ of
size $n\times m$ bits. We showed that HSRC allow this encoding by using data from the source node as well as by
using data from other storage nodes. In this subsection we describe one method to practically implement HSRC to
encode larger data objects of size $M$, where $M>k\times m$.

The first step to encode an object of size $M$ is to split it into
$u=M/(k\times m)$ vectors\footnote{We assume that the object size $M$ is
multiple of $k\times m$, otherwise the object can be encoded by parts and/or
zero-padded to meet this requirement.} of size $k\times m$ bits. Let us
represent the object to be encoded as $\bv o=(o_1,\dots,o_M)$.  After the
splitting process, $\bv o=(\bar{\bv o}_1,\dots,\bar{\bv o}_u)$, where $\bar{\bv
o}_i=(o_{k(i-1)+1},\dots,o_{k(i-1)+k})$, $o_{k(i-1)+j}\in\FF_{2^m}$,
$j=1,\ldots,k$. Each of these vectors $\bar{\bv o}_i$ is individually encoded
using the polynomial~(\ref{e:poly}) to obtain an encoded vector $\bar{\bv
r}=(\bar{\bv r}_1,\dots,\bar{\bv r}_u)$, with $\bar{\bv
r}_i=(r_{i,1},\ldots,r_{i,n})$, $i=1,\ldots,u$. Finally, the vector $\bv r=(\bv
r_1,\dots,\bv r_n)$ with the $n$ fragments to be stored in the system is
obtained by concatenating individual elements of $\bar{\bv r}$ so that $\bv
r_i=(r_{1,i},\dots,r_{u,i})$, namely ${\bf r}_i$ contains those coefficients
$r_{j,i}$ of $\bar{\bf r}$ that have $i$ as second index.

\begin{example}\rm
\label{x:hsrc}
Consider the $\langle n\!\!=\!\!7,k\!\!=\!\!3\rangle$ HSRC and the object $\bv o=(o_1,\dots,o_9)$,
where $o_i\in\mathbb F_2$.  We split the object into $u=3$ vectors $\bv o=(\bar{\bv o}_1,\bar{\bv o}_2,\bar{\bv o}_3)$,
where $\bar{\bv o}_1=(o_1,o_2,o_3)$, $\bar{\bv o}_2=(o_4,o_5,o_6)$ and $\bar{\bv o}_3=(o_7,o_8,o_9)$.  After encoding
each of the individual vectors we obtain the set of redundant vectors $\bar{\bv r}=(\bar{\bv r}_1,\dots,\bar{\bv r}_3)$
($\bar{\bv o}_i$ is encoded to obtain $\bar{\bv r}_i$), where $|\bar{\bv r}_1|=|\bar{\bv r}_2|=|\bar{\bv r}_3|=7$.
Finally, we can obtain $\bv r_2=(\bar{\bv r}_{1,2},\bar{\bv r}_{2,2},\bar{\bv r}_{3,2})$, and similarly for all the
fragments.
\end{example}

\begin{remark}
Note that this encoding technique allows stream encoding. As soon as the source node receives the first $k\times m$ bits
to store,\footnote{A gateway node in a data center receiving data from a web application end user would be treated as the
source node in our model. In such scenarios, the source itself may not be in possession of the whole data in advance.}
it can generate the vector $\bar{\bv o}_1$, encode it to $\bar{\bv r}_1$, and distribute $\bar{\bv
r}_{1,1},\dots,\bar{\bv r}_{n,1}$ to the $n$ storage nodes. Similarly, when a storage node receives $\bar{\bv r}_{i,1}$
it can forward it to other nodes for in-network redundancy generation, for instance, when the source does not have
adequate bandwidth to upload all the $n$ redundant fragments.
\end{remark}

\begin{remark}
To implement computationally efficient codes one can set $m=8$, or $m=32$, for which addition can simply be done by
xoring system words, and for which efficient arithmetic libraries are available~\cite{jerasure}.
\end{remark}

In the rest of this paper we will assume that HSRC are implemented using the method described here. We will use
the term {\bf \emph{redundant fragment}} to refer to each of the redundant elements $\bv r_1,\dots,\bv r_n$, i.e., each
node stores one redundant fragment. And similarly we will use the term {\bf \emph{redundant chunk}} to refer to each of
the sub-elements $\bar{\bv r}_{1,i},\dots,\bar{\bv r}_{u,i}$ stored in each node $i$, i.e., each node $i$ can store up
to $u$ redundant chunks.

\section{Scheduling the In-Network Redundancy Generation}
\label{s:scheduling}

In-network redundancy generation has the potential to speedup the insertion of new data in distributed storage systems.
However, the magnitude of actual benefit depends on two factors: (i) the availability pattern of the source and storage
nodes, which determines the achievable throughput, and (ii) the specific schedule of data transfer among nodes subject to
the constraints of resource availability, which determines the actual achieved throughput for data backup. In this
section, we explore the scheduling problem, demonstrating that finding an optimal schedule is computationally very
expensive even with a few simplifying assumptions, and accordingly motivate some heuristics instead.

Let $s$ be a source node aiming to store a new data object to $n$ different storage nodes, and let $i$, $i=1,\dots, n$,
represent each of these $n$ storage nodes. We model our system using discrete time steps of duration $\tau$, where at
each time step nodes can be available or unavailable to send/receive redundant data. The binary variable
$a(i,t)\in\{0,1\}$ denotes this availability for each node $i$ for the corresponding time step $t$. Using this binary
variable we can define the maximum amount of data that node $i$ can upload during time step $t$ \emph{upload
capacity}) by $$u(i,t)=a(i,t)\cdot\omega_i\!\uparrow(t)\cdot\tau,$$ where $\omega_i\!\uparrow(t)$ is the upload bandwidth of node $i$ during time step $t$. Similarly, the amount of data each node can download during time
step~$t$ (\emph{download capacity}) is given by $$d(i,t)=a(i,t)\cdot\omega_i\!\downarrow(t)\cdot\tau,$$ where
$\omega_i\!\downarrow(t)$ represents the download bandwidth of node $i$ during time step $t$.

Then, we define the \emph{in-network redundancy generation network} as a weighted temporal directed graph
$G=(E(t),V(t))$, $t\geq 0$, with the set of nodes $V(t)\subset \{s,1,2,\dots,n\}$, and the set of edges
$E(t)=\{(i,j)|i,j\in V(t)\}$.  The amount of data that nodes might send among themselves is a mapping $f:E(t)\rightarrow
\mathbb R^+$, denoted by $f(i,j,t)$, $\forall (i,j)\in E(t)$, $t\geq0$.

HSRC characteristics constrain the mapping $f$ since nodes can only send or
receive data trough valid redundancy creation triplets:
\begin{equation}
\begin{array}{c}
\not\exists ~c\in\mathcal C \text{ s.t. } c=(i,j)\vdash k \Rightarrow f(i,k,t)=0.\\
\exists~c\in\mathcal C \text{ s.t. } c=(i,j)\vdash k \Rightarrow f(i,k,t) \geq 0.
\end{array}
\label{e:c:triplets}
\end{equation}
Furthermore, we assume (for algorithmic simplicity) that nodes send data through each of the redundancy generation
triplets symmetrically:
\begin{equation}
R(c,t)=f(i,k,t)=f(j,k,t),~\forall~c\in\mathcal C;~c=(i,j)\rt k.
\label{e:c:symmetry}
\end{equation}
For ease of notation we will refer to the data sent
through each of the redundancy generation triplets simply by $R(c,t)$.

Similarly, because of the upload/download bandwidth constraints, the mapping $f$ must also satisfy the following
constraints:
\begin{itemize}
\item The amount of data the source uploads is constrained by its upload capacity:
\begin{equation}
\sum_{i=1}^n f(s,i,t)\leq u(s,t);~\forall i \in V(t).
\label{e:c:uploadsrc}
\end{equation}
\item The amount of data storage nodes upload is also constrained by their upload capacity:
\begin{equation}
\sum_{\substack{c\in O(i)}} \!\!\!R(c,t)\leq u(i,t);~\forall i \in V(t).
\label{e:c:uploadbw}
\end{equation}
\item The amount of data storage nodes download is restricted by their download capacity:
\begin{equation}
f(s,i,t)+ 2\!\!\!\sum_{\substack{c\in I(i)}} \!\!\!R(c,t)\leq d(i,t); ~\forall i \in V(t).
\label{e:c:downloadbw}
\end{equation}
\end{itemize}

A {\em Bandwidth-Valid In-Network Redundancy Generation Scheduling} is any mapping $f$ on $G$ that satisfies the
constraints defined in equations~(\ref{e:c:triplets}), (\ref{e:c:symmetry}), (\ref{e:c:uploadsrc}), (\ref{e:c:uploadbw})
and (\ref{e:c:downloadbw}).

\subsection{Optimal Schedule}
\label{s:optimal}

Let $\theta(i,t)$ be the amount of data that node $i$ had received at the end of time step $t$. For sufficiently large
enough files and small values of $m$ (e.g. $m=8$), we can assume without loss of generality that $\theta(i,t)/m$
corresponds to the index of the last {\em redundant chunk} received by node $i$. Let $\widehat M(\bar t)$ denote the
size of the largest possible file that a schedule $f$ can store in $\bar t$ time steps, which represents a data
insertion throughput of $\widehat M(\bar t)/(\bar t\tau)$. Then, by definition of erasure codes, to consider that a
file of size $\widehat M(\bar t)$ has been successfully stored after $\bar t$ time steps, each node must receive an
amount of data equal to $\widehat M(\bar t)/k$. Using this fact we can define $\widehat M(\bar t)$ as:
\begin{equation}
\widehat M(\bar t) = \text{min}\left( \theta(1,\bar t),\dots,\theta(n,\bar t)\right)\times mk.
\label{e:throughput}
\end{equation}

For a given network $G$ and a duration $\bar t$, an in-network redundancy generation scheduling $f$ is then
\emph{optimal} if it maximizes $\widehat M(\bar t)$. Note that to maximize
$\widehat M(\bar t)$ an optimal schedule will tend to even out the amount of data $\theta(i,\bar t)$ sent to each
node, obtaining a maximum insertion throughput and a minimum consumption of network resources when $\theta(1,\bar
t)=\dots=\theta(n,\bar t)$. We can measure the overall network traffic required by any schedule $f$, after $\bar t$ time steps, namely $T(f,\bar t)$, by:
\begin{equation}
T(f,\bar t) = \sum_{t=0}^{\bar t} \left( \sum_{i=1}^n f(s,i,t) + 2\sum_{c\in\mathcal C}R(c,t) \right).
\label{e:innettraf}
\end{equation}
Accordingly, we define an in-network redundancy generation schedule $f$ to be an {\em optimal minimum-traffic} schedule
if besides maximizing $\widehat M(\bar t)$, it also minimizes $T(f,\bar t)$.

From (\ref{e:innettraf}) we also want to note that in the total traffic required by the schedule $f$,
$T(f,\bar t)$, the amount of redundant data each node $i$ receives comes from two distinct components (the two summands
within parentheses): $f(s,i,t)$ and $R(c,t)$, where $c=(j',j'')\rt i$. The first component represents the redundant data
inserted by the source while the second component represents the redundant data created through the in-network
redundancy generation process. Since in the last case there are two nodes (i.e., $j'$ and $j''$) uploading data to node
$i$ the value has two be multiplied by two, leading to the following remark:

\begin{remark}
The new redundant data created through the in-network redundancy generation process requires twice the traffic
required by the source redundancy generation.
\label{r:sourcetraffic}
\end{remark}

However, although the use of the in-network redundancy generation increases the required network traffic, it has the
potential to increase the opportunities of redundancy generation, and hence, increase the data insertion throughput. In
Section~\ref{s:eval} we will show that the relative increase of insertion throughput surpass the relative increase of
network traffic, demonstrating the practicality of the in-network redundancy generation process and its scalability.

\subsection{Additional Scheduling Constraints}

\begin{figure}
\centering
\includegraphics[width=\textwidth]{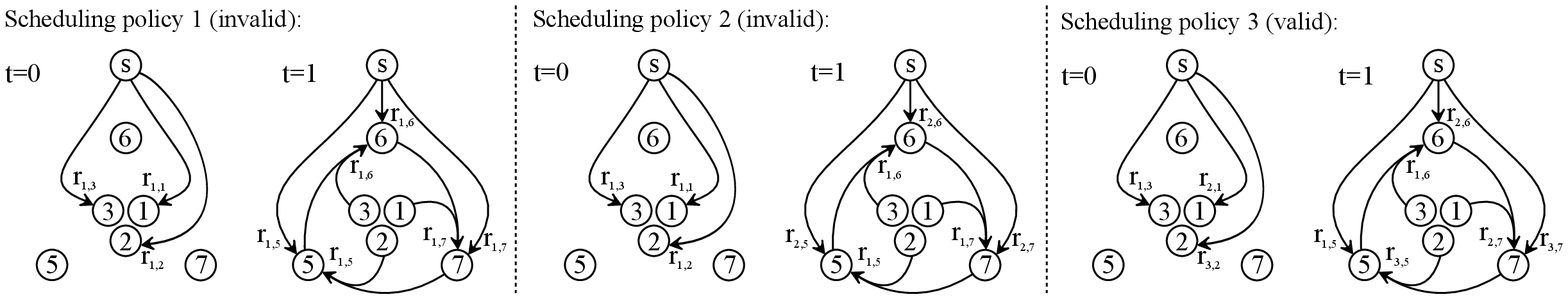}
\caption{Example of 3 different in-network redundancy scheduling policies for a system where the source node can only
upload data concurrently to 3 different nodes. Although the 3 different schedules satisfy the constraints, only
scheduling policy 3 is valid.}
\label{f:onions}
\end{figure}

In this section we elaborate that while being a bandwidth-valid schedule is a necessary condition, it is not a
sufficient condition for the schedule to be actually valid. For that, we will use Example~\ref{x:hsrc}, an in-network
redundancy generation network using HSRC with parameters $\langle n\!\!=\!\!7,k\!\!=\!\!3\rangle$, where the
redundant fragments $\bv r_1,\dots,\bv r_7$ have to be stored in nodes $1,\dots,7$ respectively.  Recall also that each
redundant fragment $\bv r_i$ is composed of 3 redundant chunks, hence $|\bv r_i|=3$.  For ease of notation we will
assume that each redundancy generation triplet $c=(i,j)\rt k$, $c\in\mathcal C$, satisfies the property $k=i\oplus j$
where $\oplus$ denotes the \emph{bitwise xor operation}. Based on Example~\ref{x:hsrc}, we consider three different
scheduling policies, all depicted in Figure~\ref{f:onions}. We assume that due to the limited upload capacity of the
source node it can only upload three redundant fragments simultaneously.

In the first scenario, at time $t=0$ the source node sends to nodes 1, 2 and 3 their first redundant chunk; at time time
$t=1$ it does the same for nodes 5, 6 and 7. Note that if at time step $t=1$ the mapping $f$ tries to make use of the
in-network redundancy generation triplets $(1,6)\rt7$, $(2,7)\rt5$ and $(3,5)\rt6$; nodes 5, 6 and 7 end up receiving
the same redundant fragment twice. In this case the in-network redundancy traffic does not contribute in speeding up the
backup process and only consumes communication resources. Although avoiding this problem is implicit in the definition
of a {\em minimum-traffic} scheduling, it needs to be explicitly considered during the scheduling.

Consider a second scheduling policy trying to solve the previous problem by sending to nodes 5, 6 and 7 the second chunk
instead of the first. It allows these nodes to receive two different fragments by time $t=1$. However, it appears a
circular dependency problem with triplets $(1,6)\rt7$, $(2,7)\rt5$ and $(3,5)\rt6$. To show this dependency, imagine
that we want to generate fragment $\bv r_{6,1}$ using non-source data. Note that $\bv r_{6,1}$ requires of
$\bv r_{5,1}$, and $\bv r_{5,1}$ requires of $\bv r_{7,1}$, which at the same time requires
of the fragment we aim to generate, $\bv r_{6,1}$.  Although it is a {\em bandwidth-valid} schedule, the
circular dependency problem makes it an unfeasible schedule.

Finally, in the third case we see how the circular dependency problems can be avoided if the source sends uncorrelated
fragments at each time step. It is easy to see from this example that a valid schedule needs to be not only
bandwidth-valid, but also ensure that: (i) nodes do not receive duplicated data, and (ii) circular triplet dependencies
are prevented.

\subsection{Complexity Analysis}

We show that finding an optimal schedule satisfying all the previous requirements is computationally very expensive, even under further simplifying assumptions:

\begin{assumption}
The amount of data that the source node $s$ sends during each time step $t$ to any storage node $i$, $f(s,i,t)$, is a
constant value and is not part of the optimization problem.
\label{a:simpleprob1}
\end{assumption}

\begin{assumption}
Storage nodes can only receive redundant chunks sequentially. It means that node $i$ will never receive chunk $\bv
r_{j+1,i}$ before previously receiving chunk $\bv r_{j,i}$.
\label{a:simpleprob2}
\end{assumption}

It is easy to see that the simplified problem subject to these two assumptions corresponds to a specific instance of the
generic case described above. The interesting property about this simplified version of the problem is that we can
reduce the decision of choosing the optimal schedule $f$ to an algorithm ``SortedVector'' which sorts $\mathcal C$, as
it is shown in Algorithm~\ref{a:algo}. It is also easy to see, how due to the iterative use of redundancy generation
triplets, Algorithm~\ref{a:algo} avoids both the ``duplicate data'' and the ``circular dependencies'' problems. However,
since $|\mathcal C|=n(n-1)$, it means that there are $(n(n-1))!$ possible ways of sorting $\mathcal C$, and thus, $\bar
t\times(n(n-1))!$ different scheduling possibilities. Thus, a brute force algorithm to determine the best schedule would
have a $O(n!)$ cost.

\begin{algorithm}[t]
\footnotesize
\caption{Creating a valid optimal schedule under Assumption~\ref{a:simpleprob1} \& Assumption~\ref{a:simpleprob2}.}
\label{a:algo}
\begin{algorithmic}
\FOR{$i$ {\bf in} $1,2,\dots,n$}
    \STATE $\theta(\text{node},0) \gets 0$
\ENDFOR
\FOR{$t$ {\bf in} $0,1,2,\dots,\bar t$}
    \FOR{$i$ {\bf in} $1,2,\dots,n$}
        \STATE $\theta(i,t) \gets \theta(i,t)+f(s,i,t)$
    \ENDFOR
    \STATE $\text{triplets} \gets \text{SortedVector}(\mathcal C)$
    \FOR{$z$ {\bf in} $1,\dots,|\mathcal C|$}
        \STATE $c \gets \text{triplets}[z]$
        \STATE /* $c=(i,j)\rt k$ */
        \STATE availDataByBw $\gets$ min($u(i,t),~u(j,t),~d(k,t)$)
        \STATE availDataByIndex $\gets$ min($\theta(i,t),~\theta(j,t))-\theta(k,t)$
        \STATE availDataByIndex $\gets$ max(availDataByIndex,~0)
        \STATE availData = min(availDataByBw,~availDataByIndex)
        \STATE $R(c,t) \gets$ availData
        \STATE $\theta(k,t) \gets \theta(k,t)~+$ availData
    \ENDFOR
\ENDFOR
\end{algorithmic}
\normalsize
\end{algorithm}

\begin{figure}
\centering
\includegraphics[scale=1]{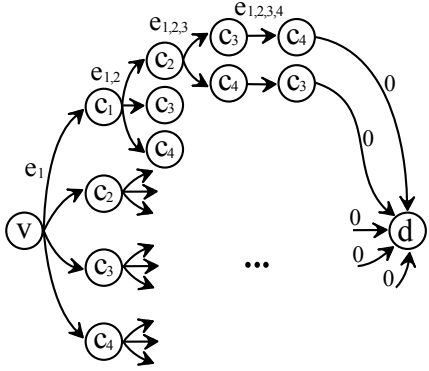}
\caption{Example of a permutation tree to implement ``SortedVector'' in Algorithm~\ref{a:algo}. We assume in this
example a hypothetical system where $|\mathcal C|=4$.}
\label{f:permtree}
\end{figure}

If we focus on a single time step $t$, then the scheduling problem can be restated as how to choose the best permutation
of $\mathcal C$. We can represent this decision problem using a permutation tree as is depicted in
Figure~\ref{f:permtree}.  The weight of the edges in this permutation tree correspond to the negative amount that
choosing each edge contributes to $\widehat M(t)$. Choosing the best scheduling algorithm tis the same than finding the
shortest path between vertices $v$ and $d$ in the permutation tree. The Bellman-Ford algorithm can find the shortest
past with cost $O(|E|\cdot|V|)$ where $|E|$ and $|V|$ respectively represent the number of edges and vertices in the
permutation tree.  However, in our permutation tree the number of edges and vertices are both $(n(n-1))!$, which makes
finding the optimal schedule for even the simplified problem computationally exorbitantly expensive, even for small
number of nodes $n$.  Hence, we consider the general problem described in~\ref{s:optimal} to be also intractable.

\section{Heuristic Scheduling Algorithms}
\label{s:algorithm}

In this section we investigate several heuristics for scheduling the in-network redundancy generation. We split the
scheduling problem into two parts, following the strategy presented in Algorithm~\ref{a:algo}.

The heuristics do not require Assumption~\ref{a:simpleprob1}, thus allowing the source node to send different
amounts of data to each storage node. We however still rely on Assumption~\ref{a:simpleprob2}, which allows us to model
the decision problem with a sorting algorithm, as previously outlined in Algorithm~\ref{a:algo}. Thus, the overall
scheduling problem is decomposed into the following two decisions: (i) How does the source node schedule its uploads?
(ii) How are redundancy generation triplets sorted?

\subsection{Scheduling traffic from the source}

Recall that generating redundancy directly from the source node involves less bandwidth than doing it with in-network
techniques (Remark~\ref{r:sourcetraffic}). Thus, a good source traffic scheduling should aim at maximizing the source's
upload capacity utilization. Furthermore, the schedule must also try to ensure that the source injected data can be
further used for the in-network redundancy generation.

Given a $\langle n,k\rangle$ HSRC, where $n=2^k-1$, any subset of $k$ linearly independent encoded fragments forms
a basis, denoted by $B$ (see Example \ref{ex:basis} for an illustration). Let $\mathcal B$ be the set of all the
possible bases $B$. Since each storage node stores one redundant fragment, we use $\widehat{\mathcal B}(t)$ to represent
all the basis of $\mathcal B$ whose corresponding storage nodes are available at a time step $t$ (and likewise, refer to
each combination of such nodes as an \emph{available basis}): $$ \widehat{\mathcal B}(t)= \left\{ B\in\mathcal B~|~
a(i,t)=1,~~\forall i\in B \right\}.  $$

From the set of available basis, $\widehat{\mathcal B}(t)$, the source node
selects one basis $B$ and uploads some data to each node $i\in B$. The amount
of data the source uploads to each node $i\in B$ is set to guarantee that at
the end of time step $t$, all these nodes have received the same amount of
data, $\theta(i,t)=\theta(j,t),~\forall i,j\in B$.  From
equations~(\ref{e:throughput}) and~(\ref{e:innettraf}) we know that evening out
the data all nodes receives allows to minimize network traffic and maximize
insertion throughput. To even out the amount of data each node in basis
receives and maximizing the utilization of the upload
capacity\footnote{We assume that the upload capacity of the source is less
than the download capacity of the basis nodes.} of the source, the source needs
to send to each node $i\in B$ an amount of data equal to
$$\frac{1}{|B|}\left(u(s,t) + \sum_{j\in B}\theta(j,t-1)\right)-\theta(i,t).$$

Besides determining how to distribute the upload capacity of the source between nodes of a basis $B$, the source
node also needs to select the basis $B$ from all the available ones $\widehat{\mathcal B}(t)$ (if more than one basis is
available). We consider the following heuristic policies for the source node to select a specific basis $B$:
\begin{itemize}
\item {\bf Random:} $B$ is randomly selected from $\mathcal B(t)$.  Repeating this procedure for several time steps is
expected to ensure that all nodes receive approximately the same amount of data from the source.

\item {\bf Minimum Data:} The source selects the basis $B$ that on an average has received less redundant data. It means
that $B$ is the basis that minimizes $\frac{1}{k}\sum_{i\in B}\theta(i,t)$. This policy tries to homogenize the amount
of data all nodes receive.

\item {\bf Maximum Data:} The source selects the basis $B$ that on an average has received more redundant data. It means
that $B$ is the basis that maximizes $\frac{1}{k}\sum_{i\in B}\theta(i,t)$. This policy tries to have a basis of nodes
with enough data to allow the in-network redundancy generation for the entire data object even when the source may not
be available.

\item {\bf No Basis:} The source does not considers any basis and instead uploads data to all the online nodes. The upload
bandwidth of the source is also distributed to guarantee that, after time step $t$, all online nodes have received the
same amount of data.
\end{itemize}

\subsection{Sorting the redundancy generation triplets}

At each time step $t$, once the source allocates its upload capacity to nodes
from a specific available basis $B$, the remaining upload/download capacity of
the available nodes is used for in-network redundancy generation. For that
purpose, the list of \emph{available triplets}, $\widehat{\mathcal C}(t)$, is
determined as follows: $$\widehat{\mathcal C}(t)= \left\{ (i,j)\rt
k~\in\mathcal C~|~ a(i,t)=a(j,t)=a(k,t)=1,~~\forall (i,j)\rt k~\in B
\right\}.$$ Then, the set of available triplets $\widehat{\mathcal C}(t)$ is
sorted, and the available upload/download capacity of storage nodes allocated
according to this priority (i.e., the first available triplets have more
preference). We consider the following sorting heuristics:

\begin{itemize}
\item {\bf Random:} Repair triplets are randomly sorted. This policy tries to uniformly distribute the utilization of
network resources to maximize the amount of in-network generated data.

\item {\bf Minimum Data:} The list of available triplets are sorted in ascending order according to the amount of data
$\theta(k,t)$ the destination\footnote{Node $k$ is the destination of a triplet $c$, $c=(i,j)\rt k$.} node $k$ has
received. This policy tries to prioritize the redundancy generation in those nodes that have received less redundant
data.

\item {\bf Maximum Data:} Similarly to the {\em Minimum Data} policy, however, triplets are sorted in descending order.
This policy tries to maximize the amount of data some specific subset of nodes receive, to allow them to sustain the
redundancy generation process even when the source is not available.

\item {\bf Maximum Flow:} The triplets are sorted in descending order according to the amount of redundant data these
nodes can help generate. Note that the amount of data a triplet $c$ can generate at each time step $t$, where
$c=(i,j)\rt k$, is given by:
\begin{align*}\text{min}(&u(i,t),~u(j,t),~d(k,t),\\&\theta(i,t)-\theta(k,t),\\&\theta(j,t)-\theta(k,t))\end{align*} This
policy tries to maximize the amount of new redundancy generated per time step.
\end{itemize}

\section{Experimental results}
\label{s:eval}

\begin{table}
\centering
\begin{tabular}{ccc}\toprule
Policy Name & Source Policy & In-Network Policy \\ \toprule
{\bf RndFlw} & random & maximum flow \\ \midrule
{\bf RndDta} & random & minimum data \\ \midrule
{\bf MinFlw} & minimum data & maximum flow \\ \midrule
{\bf MinDta} & minimum data & minimum data \\ \bottomrule
\end{tabular}
\caption{Different policy combinations.}
\label{t:policies}
\end{table}

He have proposed four different policies for the source traffic scheduling
problem and four policies for the triplets sorting problem. However, after an
extensive experimental evaluation of all polices we will only report for each
case the two best policies (in terms of achieved throughput). At the source,
the {\em random} and {\em minimum data} policies consistently outperform the
others, and at the storage nodes, the {\em maximum flow} and {\em minimum data}
sorting policies for the triplets likewise outperform the others.  We will
refer to each of the combinations as denoted in Table~\ref{t:policies}.
First, the interpretation of the good performance of the random policy in
the source node is that the use of random bases favors the diversity of
information among the nodes, which in turn enables more redundancy generation
triplets. Second, it is interesting to note that the {\em minimum data} policy
obtains good storage throughput in both cases, which leads us to infer that
{\em in general, prioritizing redundancy generation in those nodes that have
received less data} is a good strategy to maximize the throughput of the backup
process.

\subsection{Setting}

We considered a $\langle n\!\!=\!\!7,k\!\!=\!\!3\rangle$-HSRC, which is a code that can achieve a static data
resiliency similar to a 3-way replication, but requiring only a redundancy factor of $7/3\simeq2.33$.\cite{OD} Using
this erasure code we simulated various backup processes with different node (un)availability patterns for a fixed number
of time steps $\bar t$. In all the simulated cases we consider three different metrics:
\begin{enumerate}[(i)]
\item The maximum amount of data that can be stored in $\bar t$ time steps, $\widehat{M}(\bar t)$.
\item The amount of data the source node uploads per
unit of useful data backed up, $$\frac{1}{\widehat{M}(\bar t)}\sum_{t=0}^{\bar t}\sum_{i=1}^n f(s,i,t).$$
\item The total traffic generated per unit of useful data stored, $T(f,\bar t)/\widehat{M}(\bar t)$.
\end{enumerate}

We evaluate the three metrics for a system using an in-network redundancy generation algorithm and we compare our
results with a system using the naive erasure coding backup process, where the source uploads all the data directly to
each storage node. Our results depict the savings and gains, in percentage, of using an in-network redundancy algorithm
with respect to the naive approach.

Regarding the (un)availability patterns of nodes and their bandwidth constraints we consider two different distributed
storage cases:
\begin{enumerate}[(i)]
\item A P2P-like environment where we assume, to simplify simulations, that nodes have an upload
bandwidth uniformly distributed between 20Kbps and 200Kbps, and an asymmetric
download bandwidth equal to four times their upload bandwidth. Nodes in this
category follow two different availability traces from real decentralized
application: (i) traces from users of an instant messaging (IM)
service~\cite{f2favail} and traces from P2P nodes in the aMule KAD DHT
overlay~\cite{globalkad}. In both cases we filter the nodes that on average
stay online more than 4, 6 and 12 daily hours, obtaining different mean
availability
scenarios.
\item Real availability traces collected from a Google
datacenter\footnote{Publicly available at: http://code.google.com/p/googleclusterdata/}. The traces contain the
normalized I/O load of more than 12,000 servers monitored for a period of one month. We consider that a server is
available to upload/download data when its I/O load is under the $p$-percentile load. We consider three different
percentiles, $p=0.25,0.5,0.75$, giving us three different node availability constraints.
\end{enumerate}
Finally the time step duration is set to $\tau=1$hour and we obtain the results by averaging the results of
500 backup processes of $\bar t=120$ time steps each (5 days).

Before discussing the results, we will like to note that the experiments make a few simplifying assumptions. Furthermore, real deployments have somewhat different workload characteristics than what have been considered above. Hence, the quantitative results we report are only indicative (and many more settings could possibly be experimented), and instead the specific choices help us showcase the potential benefits of our approach in only a qualitative manner.

\subsection{Results}

\begin{figure}[htbp]
\centering
\subfloat[IM traces.]{\includegraphics[scale=0.9]{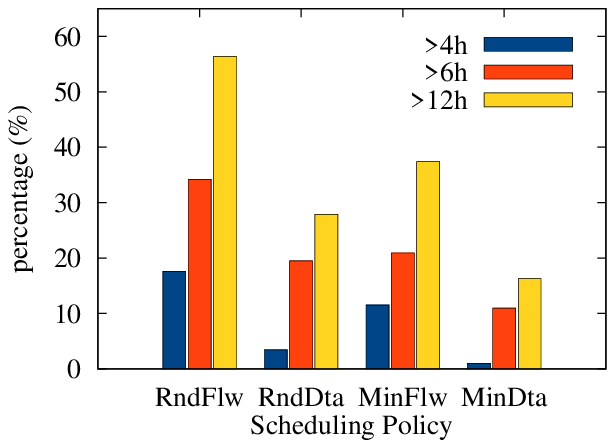}}
\subfloat[P2P traces.]{\includegraphics[scale=0.9]{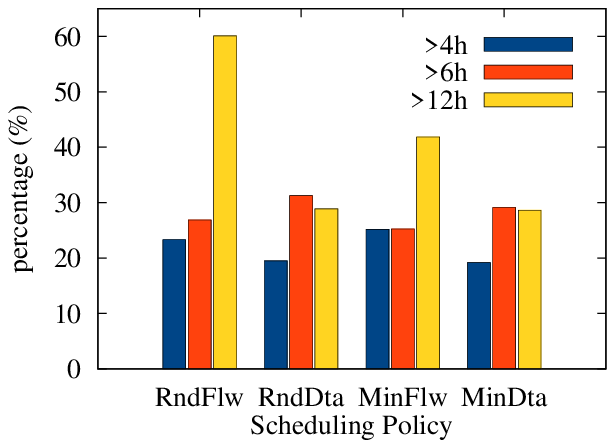}}
\subfloat[Google traces.]{\includegraphics[scale=0.9]{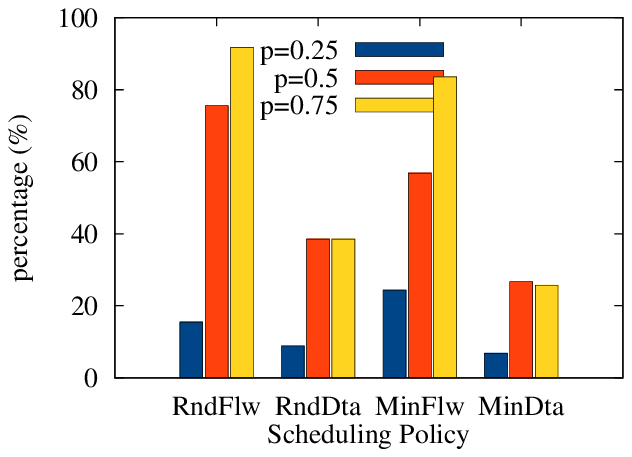}}
\caption{Increment of the maximum amount of stored data (throughput) for the three different availability traces.}
\label{f:stored}
\end{figure}

\paragraph{Storage Throughput.} In Figure \ref{f:stored} we show the increment of the data insertion throughput achieved
by the in-network redundancy generation process.  We can see how the gain is higher when nodes are more available for
redundancy generation.  This fact is a consequence of the constraint in eq.~(\ref{e:c:symmetry}) requiring redundancy
generation triplets to be symmetric, which requires the three involved nodes in each triplet to be available
simultaneously. The higher the online availability, the higher the chances to find online three nodes from a triplet.
Further, we observe that the {\em RndFlw} policy achieves significantly better results in comparison to other policies;
the second best policy is {\em MinFlw}. It is easy then to see that the {\em Maximum Flow} heuristic plays an important
role on the overall redundancy generation throughput, which tries to maximize the use of those nodes that can
potentially generate more redundancy. Additionally, a {\em Random} source selection policy provides more benefits than
the {\em Minimum Data} policy.

\begin{figure}[htbp]
\centering
\subfloat[IM traces.]{\includegraphics[scale=0.9]{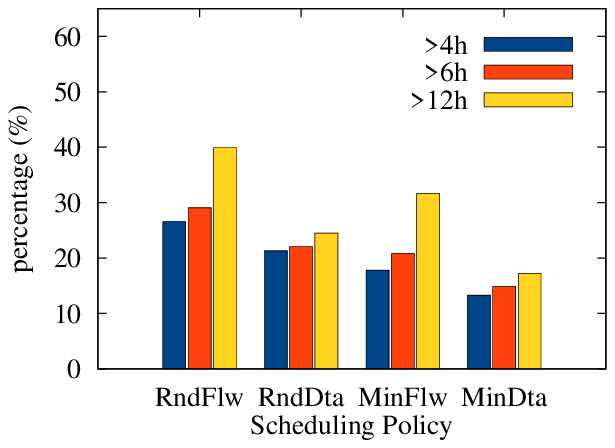}}
\subfloat[P2P traces.]{\includegraphics[scale=0.9]{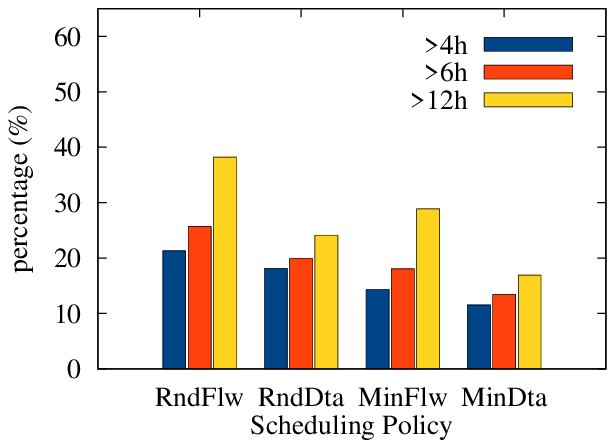}}
\subfloat[Google traces.]{\includegraphics[scale=0.9]{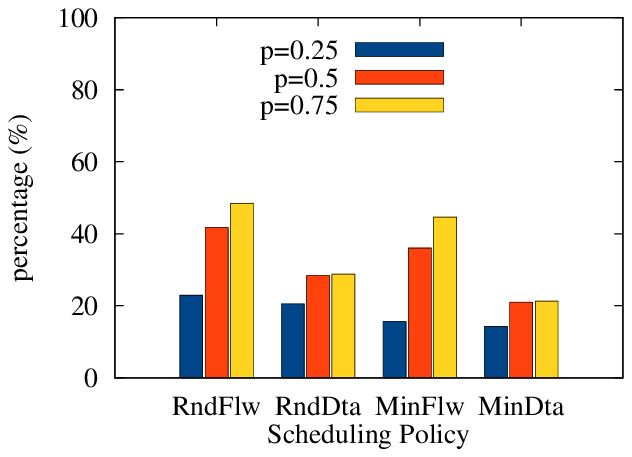}}
\caption{Increment of the required network traffic of the in-network redundancy generation strategy for the three different availability traces.}
\label{f:traffic}
\end{figure}

\begin{figure}[htbp]
\centering
\subfloat[IM traces.]{\includegraphics[scale=0.9]{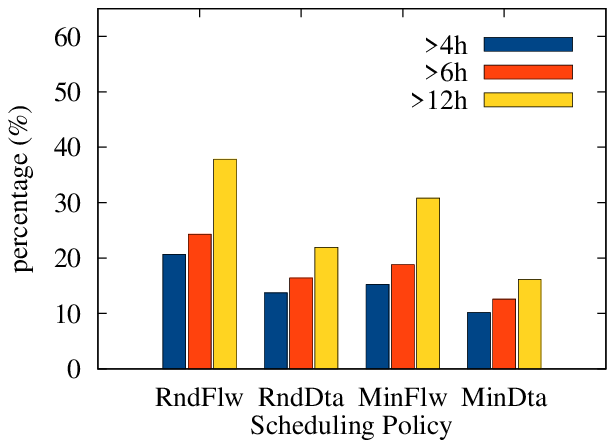}}
\subfloat[P2P traces.]{\includegraphics[scale=0.9]{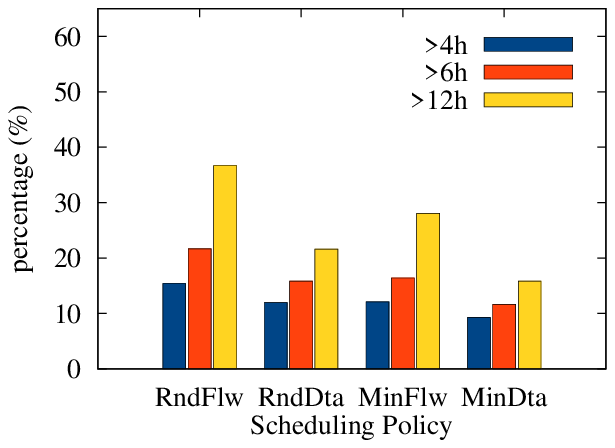}}
\subfloat[Google traces.]{\includegraphics[scale=0.9]{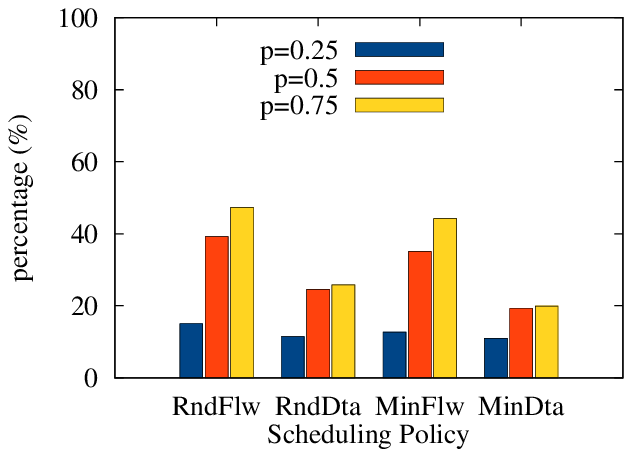}}
\caption{Reduction of the data uploaded by the source for the three different availability traces.}
\label{f:source}
\end{figure}

\paragraph{Network Traffic.} In Figure~\ref{f:traffic} we show the increment
on the required network traffic of the in-network redundancy generation
strategy as compared to the traditional redundancy generation. As noted
previously (in Remark~\ref{r:sourcetraffic}), the total traffic required for
in-network redundancy generation can be up to twice the needed by the
traditional process (i.e. 100\% traffic increment). However, since the
in-network redundancy generation cannot always take place due to the node
availability constraints, the traffic increment is always below 100\%. As it
is expected then, the traffic increment is minimized when nodes are less
available, in which case the source has to generate and introduce larger
amounts of redundancy (i.e., less reduction in the data uploaded by source,
as shown in Figure~\ref{f:source}). It is also important to note that the
increase in traffic is approximately the same or even less than the increase
in storage throughput even for low availability scenarios. Thus the
in-network redundancy generation scales well by achieving a better
utilization of the available network resources than the classical storage
process.

\paragraph{Data Uploaded by the Source.} In Figure~\ref{f:source} we show the reduction of data uploaded by the source.
In the traditional approach, the source needs to upload $7/3\simeq2.33$ times the size of the actual data to be stored;
$4/7 \simeq 57\%$ of this data is redundant, however the in-network redundancy generation process allows to reduce the
amount of data uploaded by the source. In this figure we can see how in the best case ({\em RndFlw} policy) our approach
reduces the source's load by 40\% (out of a possible 57\%), yielding 40-60\% increase in storage throughput.

Finally, we want to note that the in-network redundancy performance requires finding three available nodes
simultaneously, which becomes difficult on environments with fewer backup opportunities. To solve this problem, we would
need to look at more sophisticated in-network redundancy generation strategies not subjected to the symmetric constraint
(defined in eq.~(\ref{e:c:symmetry})), so that nodes can forward and store partially-generated data. However, the
scheduling problem will be much more complicated, and is beyond the reach of this first work. Furthermore, in real
traces, nodes will have correlation (e.g., based on batch jobs), which are missing in the synthetic traces, and such
correlations can be leveraged in practice. Exploring both these aspects will be part of our future work.


\section{Conclusions}
\label{s:conclusions}
In this work we propose and explore how storage nodes can collaborate among themselves to generate erasure encoded redundancy by leveraging novel erasure codes' local-repairability property. Doing so not only reduces a source node's load to insert erasure encoded data, but also significantly improves
the overall throughput of the data insertion process. We demonstrate the idea using self-repairing codes. We show that determining an optimal schedule among nodes to carry out in-network redundancy generation subject to resource constraints of the system (nodes and network) is computationally prohibitive even
under simplifying assumptions. However, experiments supported by real availability traces from a Google data center, and P2P/F2F applications show that some heuristics we propose yield significant gain in storage throughput under these diverse settings, proving the practicality of not only the idea in general, but also that of the specific proposed heuristics.

\end{document}